\documentclass[pdflatex]{sn-jnl}

\usepackage[compatibility=false]{caption}
\usepackage{graphicx}
\usepackage{multirow}
\usepackage{amsmath,amssymb,amsfonts}
\usepackage{amsthm}
\usepackage{mathrsfs}
\usepackage[title]{appendix}
\usepackage{xcolor}
\usepackage{textcomp}
\usepackage{manyfoot}
\usepackage{makecell}
\usepackage{booktabs}
\usepackage{algorithm}
\usepackage{algorithmicx}
\usepackage{algpseudocode}
\usepackage{listings}
\usepackage{subcaption}
\usepackage{multirow}
\usepackage{lineno}
\usepackage{tikz}
\usetikzlibrary{shapes,arrows}
\tikzstyle{startstop} = [ellipse, minimum width=5cm, minimum height=1cm, text centered, draw=black]
\tikzstyle{process} = [rectangle, minimum width=8cm, minimum height=1cm, text centered, draw=black]
\tikzstyle{arrow} = [thick,->,>=stealth]
\usepackage{comment}
\usepackage[separate-uncertainty=true,list-units=repeat,multi-part-units=brackets]{siunitx}
\usepackage{lmodern}
\usepackage{anyfontsize}
\usepackage{natbib}
\setcitestyle{aysep={}} 

\usepackage{lineno}
\usepackage{tabularx}
\usepackage{tikz}
\usetikzlibrary{shapes.geometric, arrows.meta, positioning, calc}
\tikzstyle{block} = [
    rectangle,
    rounded corners=2pt,
    minimum width=7.2cm,     
    minimum height=0.5cm,
    text centered,
    draw=black,
    fill=white,
    text=black,
    font=\footnotesize
]
\tikzstyle{arrow} = [thick, ->, >=stealth, draw=black]
\tikzstyle{note} = [font=\scriptsize\itshape\color{black}]
\usepackage{multirow}

\theoremstyle{thmstyleone}%
\theoremstyle{thmstyletwo}%

\theoremstyle{thmstylethree}%

\begin{document}

\title{Orthotropic Viscoelastic Creep in Cellular Scaffolds}


\author*[1]{\fnm{Alessia} \sur{Ferrara}}\email{aferrara@ethz.ch}

\author[1]{\fnm{Falk K.} \sur{Wittel}}\email{fwittel@ethz.ch}
\equalcont{These authors contributed equally to this work.}

\affil*[1]{\orgdiv{Institute for Building Materials}, \orgname{ETH Zurich}, \orgaddress{\street{Laura-Hezner-Weg 7}, \city{Zurich}, \postcode{8093}, \country{Switzerland}} ORCID: 0009-0003-9627-5200}

\abstract{Recent measurements of Norway spruce have revealed stress-state-dependent normalized creep behavior, highlighting a gap in our fundamental understanding. This study examines whether the anisotropic response originates from the micro-structural, cellular nature of composite cell walls with varying tracheid types. Cell wall creep parameters are identified via surrogate-based inverse parameter identification, applied to hierarchical micro-mechanical and FEM models of increasing topological complexity up to the growth ring scale. Despite microstructural disorder, simulated creep curves converge toward a universal set of proportionality factors. The results indicate that directional creep behavior cannot be attributed solely to tissue-scale topology, and that realistic predictions require the inclusion of non-linear material responses at stress concentration sites.}


\maketitle

\section{Introduction}\label{sec1}
In the past few years, timber engineering has faced a tremendous acceleration in the development of mass timber structures of increasing height and spanwidth in urban settings \cite{villamizar_santamaria_high-rise_2024}. The integration of novel timber products and design principles into standardization, the increased use of hybrid structures with other species and materials, and the green aspects of wood as a carbon sink with cascade utilization are the basis of this success. All too often, designers forget that wood is a hygro-responsive natural material that exhibits time- and moisture-dependent long-term deformations under sustained loads and varying moisture conditions throughout its entire lifespan. A fundamental understanding of wood physics and creep behavior at the material scale underpins the performance and reliability of any engineered timber systems, such as cross-laminated timber (CLT), self-shaping CLT \cite{gronquist_etal_2019}, or glued-laminated timber (GLT). Long-term deformation modeling is essential in structural design for predicting long-term deflections and ensuring the serviceability and safety of timber structures throughout their lifetime. Consequently, it is no surprise that the orthotropic viscoelastic behavior of wood has recently gained increased attention \cite{maas_comprehensive_2025,ferrara_tensile_2025,bengtsson_evaluating_2023}. It is striking to realize that creep deformations can exceed elastic deformations by a factor of 2 to 3, making them a relevant deformation mode for design purposes. 

As wood is assumed to behave orthotropic with reasonable accuracy, the viscoelastic, time-dependent creep compliance must be orthotropic, as well. Until today, in rheological wood models, this orthotropy enters only via the orthotropic elastic compliance, assuming a scalar proportionality between the skleronomous and rheonomous behavior, namely elastic and the viscoelastic orthotropic compliance tensors \cite{hassani_etal_2015,fortino_3d_2009,hanhijarvi_computational_2003}. The underlying idea is that the material reacts in a similar way to instantaneous and retained loads. This assumption greatly simplifies parameter identification, as the proportionality factor can be determined from a single creep test. For compact materials with orthotropy, such as fiber-reinforced composites, this assumption is quite accurate. For cellular, porous, hierarchical materials, such as wood, however, recent experimental observations from creep tests under all possible stress states have raised strong questions \cite{maas_comprehensive_2025,bengtsson_evaluating_2023}. It is observed for Norway spruce, the main construction wood in Europe, that different groups of components of the creep compliance tensors require different sets of proportionality constants, as well as for creep under tension or compression \cite{maas_comprehensive_2025}. When thinking about the hierarchical microstructure of wood, the proportionality assumption seems to be fairly unfounded. Rather, one would expect that the arrangement of different wood cell types, such as tracheids of early (EW), transition (TW), and late (LW) wood in alternating growth rings, would significantly affect its rheonomous behavior. 

It is a question of enormous importance in wood physics, how macroscopically observed behavior emerges from the behavior at lower scales, in this case, the tissue scale. Wood tissue models play a crucial role in linking bulk wood properties with the structural organization of wood constituents in early, transition, and latewood tracheids within growth rings. Wood as a natural hierarchical material \cite{fratzl_natures_2007} is an ideal material for multi-scale models, as one benefits from a clear separation of length scales. In principle, the combination of meaningful cell-wall models with differently arranged representative microstructural tissue models is the prerequisite for realistic forecasts of the macroscopic behavior. The hygro-mechanical skleronomous orthotopic behavior has been successfully modeled using various multi-scale approaches in the past \cite{mora_etal_2019,holmberg_persson_1999,hofstetter_hierarchical_2009,qing_3d_2010,mora_etal_2019}. As cell wall layers are compact composites formed from cellulose, hemi-cellulose, and lignin with different amounts, the assumption of an identical proportionality factor between elastic and viscelastic compliance is reasonable. The central hypothesis of this work is that the observed component-wise scaling factors on the macro-scale emerge from the hierarchical organization of the composite cell wall material in a cellular solid comprising different tracheid types. In this work, we provide a comprehensive framework for understanding the time-dependent behavior of hierarchical, cellular, and thus orthotropic materials under sustained loading conditions. The material creep model parameters on the cell wall scale are identified inversely using a surrogate optimization algorithm, which is suitable for black box optimization problems. We apply models with varying topological complexity that contribute to a growth ring model, which exhibits the effect of microstructural disorder. We demonstrate that all numerically determined creep curves collapse towards a single set of proportionality factors. As the directional dependency of the viscoelastic response does not emerge from topological details on the tissue scale, we conclude that models must include non-linear material behavior at stress concentration points to make realistic predictions for all stress components.
\section{Hierarchical Model and Inverse Parameter Determination}\label{sec2}
\begin{figure}[tb]
  \centering{\includegraphics[width=1\textwidth]{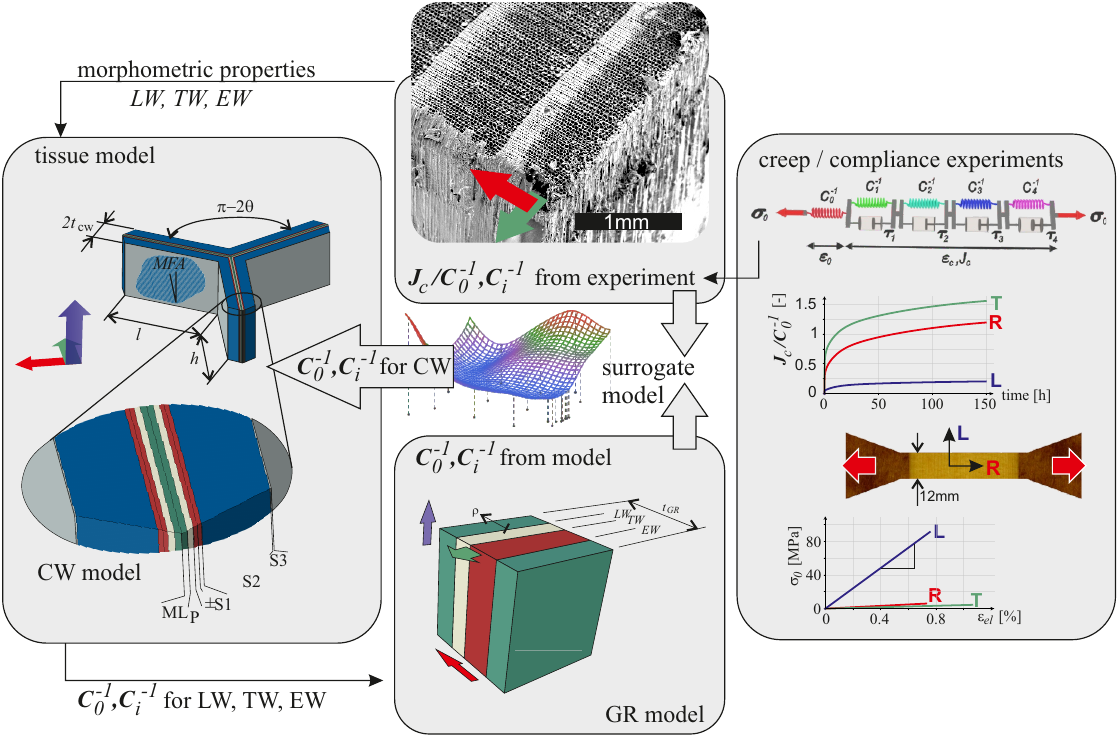}
  \caption{\label{fig:1new} Scheme of the approach taken in this work. Stress-strain and creep curves are data at $\omega$=\SI{12}{\percent} from \cite{maas_comprehensive_2025}. }}
\end{figure}

This section outlines the numerical framework used in this study, starting with the description of orthotropic viscoelasticity (Sec.~\ref{subsec2-1}). Consequently, the computation of material properties through a hierarchical multi-scale approach based on composite mixing rules and laminate theory (Sec.~\ref{subsec2-2}) is described. Section~\ref{subsec2-3} explains two alternative ways for building up tissue models for EW, TW, and LW, which provide the material behavior for the growth ring model in Sec.~\ref{subsec2-4}. Finally, Sec.~\ref{subsec2-5} explains the inverse black box parameter identification procedure based on surrogate models to find the most suitable material description. The overall procedure is outlined in Fig.~\ref{fig:1new}.
\subsection{Orthotropic viscoelasticity} \label{subsec2-1}
Orthotropy is characterized by the intrinsic symmetry of the material with respect to two perpendicular planes, reducing the compliance tensor to 9 independent parameters that need to be determined. Note that from symmetry with respect to two perpendicular planes, symmetry with respect to the third perpendicular plane follows implicitly. Clearly, for a natural material like wood, orthotropy is only a reasonable assumption that can be experimentally and numerically validated. The resulting absence of normal stress and shear deformation coupling and vice versa strongly simplifies calculations and parameter identification.
\subsubsection{Orthotropic elasticity model} \label{subsubsec2-1-1}
The orthoropic elastic compliance matrix ${\mathbf C}_{0}^{-1}$ with the \num{12} engineering parameters from which \num{9} are independent is written as
\begin{equation}
     {\mathbf C}_{0}^{-1}={\mathbf C}_{0,AB}^{-1}=\mathbf{D}_{0,AB}=\left[ \begin{array}{cccccc}
          \frac{1}{E_{1}}\;         & \;\frac{-\nu_{21}}{E_{2}}\; & \;\frac{-\nu_{31}}{E_{3}}\; & \;0\;                & \;0\; & \;0 \\
          \frac{-\nu_{12}}{E_{1}}\; & \;\frac{1}{E_{2}}\;         & \;\frac{-\nu_{32}}{E_{3}}\; & \;0\;                & \;0\; & \;0 \\
          \frac{-\nu_{13}}{E_{1}}\; & \;\frac{-\nu_{23}}{E_{2}}\; & \;\frac{1}{E_{3}}\;         & \;0\;                & \;0\; & \;0 \\ 
          \;0\;                     & \;0\;                       & \;0\;                       & \;\frac{1}{G_{23}}\; & \;0\; & \;0 \\
          \;0\;                     & \;0\;                       & \;0\;                       & \;0\; & \;\frac{1}{G_{31}}\; & \;0 \\
          \;0\;                     & \;0\;                       & \;0\;                       & \;0\; & \;0\; & \;\frac{1}{G_{12}} \\     \end{array} \right].
\end{equation}
Note that here ${\mathbf C}_{0}^{-1}$ is expressed in the Voigt notation with the indices $A,B=1\ldots 6$. Since ${\mathbf C}_{0}^{-1}$ must be symmetric ( $C^{-1}_{0,AB}=C^{-1}_{0,BA}$), the non-zero off-diagonal terms have to follow $\nu_{jk}/E_{j}=\nu_{kj}/E_{k}$ with $j,k=1,2,3$. For ${\mathbf C}_{0}^{-1}$ to be positive definite, its determinant needs to be greater than zero ($\det({\mathbf C}_{0}^{-1})>0$), all main diagonal elements have to be positive ($E_{j},G_{jk}>0$), the main subdeterminats have to be positive ($|\nu_{jk}|<\sqrt{E_{j}/E_{k}}$), and the $3\times 3$ subdeterminat needs to be positive ($\nu_{12}\nu_{23}\nu_{31}<0.5(1-\nu_{12}^2E_{2}/E_{1}-\nu_{23}^2E_{3}/E_{2}-\nu_{31}^2E_{1}/E_{3})<0.5$).

In this work, several local coordinate systems (CSys) are employed, for layers (Secs.~\ref{subsubsec2-2-1} and~\ref{subsubsec2-3-1}) and cell wall (Sec.~\ref{subsubsec2-3-2}), and one global material CSys following the anatomical directions $R-T-L$ relating to 1-2-3. The material parameters are therefore $E_R,~E_T,~ E_L,~ G_{TL}=G_{LT},~G_{LR}=G_{RL},~G_{RT}=G_{TR},~\nu_{TL}=E_T\nu_{LT}/E_L,~\nu_{RL}=E_R\nu_{LR}/E_L,~\nu_{TR}=E_T\nu_{RT}/E_R$ from which 9 are independent and need to be determined. Orthotropy can be visualized by deformation bodies in polar coordinates or the $R-T-L$ space to gain an intuitive understanding and the basis for Figs.~\ref{fig:defbodyp_polar}. In principle, this corresponds to the deformation due to a uniaxial unit load in a different orientation, or unit moment around a single differently oriented axis, respectively \citep{grimsel_mechanisches_1999}.
\subsubsection{Orthotropic Kelvin-Voigt (KV) chain-based viscoelastic model} \label{subsubsec2-1-2}
The KV series-based model effectively captures the creep behavior, as noted in previous studies \citep{gutierrez2014, hajikarimi2021}. First, its behavior is described in the framework of linear viscoelasticity. The idea behind the KV-chain is that each KV element contributes different characteristic time delays, enabling the model to capture different time spans of the material’s deformation under constant stress. In addition, this setup also allows for a straightforward extension with other moisture-dependent rheological properties \citep{hassani_etal_2015}. The total compliance $\mathbf{C}^{-1}(t)$ is composed of the elastic $\mathbf{C}_{0}^{-1}$ and creep compliance $\mathbf{J}_{c}(t)$, namely $\mathbf{C}^{-1}(t)=\mathbf{C}_{0}^{-1}+\mathbf{J}_{c}(t)$ (see Fig.~\ref{fig:1new} top right). The mathematical formulation of each component $J_{c,AB}(t)$ of the tensor ${\mathbf J}_{c}(t)$ is expressed by
\begin{equation}\label{eq:pronyser1}
J_{c,AB}(t) = \sum_{i=1}^{N}C_{i,AB}^{-1}(1-e^{-t/\tau_{i}}),
\end{equation}
where $N$ denotes the number of KV-elements, $C_{i,AB}^{-1}$ is the $AB$-component of the $i^{th}$ element compliance, and $\tau_{i}$ denotes the characteristic time of the $i^{th}$ element. In this work, the unit of hours is chosen for $t$ and $\tau$, and all compliances are in \SI{}{\per\mega\pascal}. The viscoelastic compliance tensor is assumed to be proportional to the elastic compliance tensor, which is a common assumption in literature \citep{fortino_3d_2009,hanhijarvi_computational_2003, hassani_etal_2015}, with the dimensionless scalar proportionality factors
\begin{equation}\label{eq:gamma}
\gamma_{i}^{ve} = {\mathbf C_{0}^{-1}} / {\mathbf C_{i}^{-1}}.
\end{equation}
 In this way, Eq.~\ref{eq:pronyser1} renders to
\begin{equation}\label{eq:pronyser2}
J_{c,AB}(t) = C_{0,AB}^{-1}\sum_{i=1}^{N} \frac{1}{\gamma_{i}^{ve}} (1-e^{-t/\tau_{i}}).
\end{equation}
Consistent with our previous work \citep{ferrara_tensile_2025}, we set the number of KV-elements to $N = 4$ and the characteristic times $\tau_{i},i=1\ldots 4$ \emph{a~priori} as $\tau_{i} = [0.1, 1, 10, 100]$ to ensure fitting consistency and to reduce the number of unknown parameters to $\gamma_{i}^{ve},i=1\ldots 4$.
\subsection{Micro-mechanical viscoelastic cell wall model}\label{subsec2-2}
The computation of the material properties is performed using a hierarchical multi-scale approach based on composite mixing rules and laminate theory, similar to \citet{ferrara_wittel_densification_2025}, where a detailed description can be found. Cell wall compliance tensors can be modeled by a layer-wise (LW) or an equivalent single-layer (ESL) representation with smeared material properties. When dealing with a small number of tracheids, the LW method can be applied with its discrete representation of each layer. However, for numerous tracheids, the computationally more efficient ESL representation has to be chosen, which requires only a fraction of the finite elements needed otherwise.
\subsubsection{Discrete layered cell wall model}\label{subsubsec2-2-1}
In the discrete layered cell wall model, material properties are defined at the scale of the cell wall layer, with as many orthotropic compliance tensors as bonded layers. The cell wall model shown in Fig.~\ref{fig:1new} is composed of middle lamella (ML), primary (P) wall, and three secondary (S) walls, namely S1, S2, and S3 \citep{fengel_stoll-73,holmberg_persson_1999,persson_micromechanical_2000}. The ML is characterized by a mechanically dominating amount of lignin and a 3D random distribution of hemicellulose chains, resulting in isotropic behavior. The cellulose microfibrils in the P wall are considered to be 2D randomly oriented, which results in rotational symmetry of the compliance and thus transvers isotropy around the axis perpendicular to the cell wall. For the sake of simplicity and in line with our previous work \citep{ferrara_wittel_densification_2025}, both the P wall and the ML are treated as isotropic and are assumed to have the same thickness and composition, even though P exhibits a slight transverse isotropy in spruce. The discrepancies arising from this assumption were quantified to be below $\pm 3.5\%$ and found to be insignificant. Each S-layer is assumed to be a transverse isotropic material within its layer coordinate system. Each layer has an individual cellulose microfibril angle (MFA), as well as a characteristic chemical composition (see Tab.~\ref{tab:1}). Note that the S1 layer is decomposed into two layers with positive (S1+) and negative (S1-) MFA.
\begin{table}[htbp]
\centering
\scriptsize
\renewcommand{\arraystretch}{1.1}
\caption{Ultra-structural features, chemical composition, and engineering constants of the cell wall layers. Cellulose (C), hemicelluloses (H), and lignin (L) content from \citet{persson_micromechanical_2000,qing_3d_2009}. Engineering constants are in GPa (at $\omega$=\SI{12}{\percent}) in the local layer system (1: $\parallel$ microfibrils; 2,3: $\perp$ directions).}
\label{tab:1}
\begin{tabularx}{\textwidth}{l|c c c c c|*{9}{>{\centering\arraybackslash}X}}
 & MFA & thick. & C & H & L 
 & $E_1$ & $E_2$ & $E_3$ 
 & $G_{23}$ & $G_{31}$ & $G_{12}$ 
 & $\nu_{12}$ & $\nu_{13}$ & $\nu_{23}$ \\
 & & ($\mu$m) & \multicolumn{3}{c|}{(\%)} & \multicolumn{9}{c}{(GPa / -)} \\
\hline
ML & rand. & 0.175 & 12 & 26 & 62 & 3.41 & 3.41 & 3.41 & 1.27 & 1.27 & 1.27 & 0.35 & 0.35 & 0.35 \\
P  & rand. & 0.175 & 12 & 26 & 62 & 3.41 & 3.41 & 3.41 & 1.27 & 1.27 & 1.27 & 0.35 & 0.35 & 0.35 \\
S1 & 60/-60 & 0.125 & 35 & 30 & 35 & 48.93 & 3.60 & 3.60 & 1.29 & 1.31 & 1.31 & 0.34 & 0.34 & 0.40 \\
S2 & 15 & calc. & 50 & 27 & 23 & 68.40 & 4.61 & 4.61 & 1.67 & 1.71 & 1.71 & 0.34 & 0.34 & 0.38 \\
S3 & 75 & 0.035 & 45 & 35 & 20 & 62.17 & 4.20 & 4.20 & 1.53 & 1.55 & 1.55 & 0.34 & 0.34 & 0.37 \\
\end{tabularx}
\end{table}
Since the respective thicknesses $t_{ML},t_{P},t_{S1},t_{S3}$ hardly vary across a growth ring, the thickness of the S2 layer has to vary as $t_{S2}=t_{cw}-\sum t_i$, with~$i=(ML,P,S1,S3)$ and $t_{cw}$ denoting the cell wall thickness.

The elastic material properties for each layer are calculated from the material properties of its chemical constituents, i.e, cellulose, hemicelluloses, and lignin \citep{cave:1978:1,cousins:1976,cousins:1978} combined with the respective volume fractions. The moisture dependence is incorporated via scaling factors that reduce the stiffness of lignin and hemicelluloses based on the bound water content \citep{cave:1978:1}. The composite behavior is obtained in two steps: first, by combining cellulose and hemicelluloses to form microfibrillar aggregates \citep{donaldson_2007,fernandes-etal-2011}, which is then combined with the lignin. Then, the material properties of each layer are calculated using the Halpin-Tsai equations \citep{halpin:1976}. Note that due to the moisture dependence of the lignin and hemicelluloses properties, the layer properties are also moisture dependent, as well. A detailed formulation can be found in \citet{ferrara_wittel_densification_2025}. The resulting engineering constants expressed in the layer coordinate systems at \SI{12}{\percent} moisture content ($\omega$ or m.c.) are summarized in Tab.~\ref{tab:1}. For simplicity, the moisture content is taken as constant at \SI{12}{\percent} throughout this work, which makes micromechanical calculations of hygro-expansion coefficients obsolete.

The viscoelastic compliance tensor of each layer is assumed to be proportional to the corresponding elastic compliance tensor (see Sec.~\ref{subsubsec2-1-1}) and, even though $\mathbf C_{0}^{-1}$ and $\mathbf C_{i}^{-1}$ are different, the ratios $\gamma_{i}^{ve}$ are identical for all layers. Note that this common assumption was numerically verified before this study using RVEs with elastic fibers in a viscoelastic matrix.
\subsubsection{Smeared cell wall model}\label{subsubsec2-2-2}
The proportionality assumption for the viscoelastic compliance tensor is applied identically to the ESL representation, only that multiple layers are merged into a smeared elastic compliance tensor. Following \citet{ferrara_wittel_densification_2025}, the secondary wall layers (S1, S2, S3) are merged into one ESL with an orthotropic compliance tensor. In contrast, the primary wall (P) and middle lamella (ML) form an isotropic compound middle lamella (CML) that embeds all S-ESL tracheids. The elastic properties of the layers are computed as described in Sec.~\ref{subsubsec2-2-1}. What follows is the additional calculation of the ESL compliance tensor \(\mathbf{C}_{0}^{-1,S}\) as a function of tracheid thickness using the classical laminate theory \cite{kelly_comprehensive_2000}. For the isotropic CML, we set $\mathbf{C}_{0}^{-1,CML} = \mathbf{C}_{0}^{-1,ML} = \mathbf{C}_{0}^{-1,P}$, since both layers behave very similar.
\subsection{Regular and disordered tissue models}\label{subsec2-3}
Creating representative volume elements (RVEs) for wood tissue can be quite a demanding task due to the topological variety and resulting high finite element mesh densities. In this work, two approaches are compared: (1) a Y-shaped repetitive unit cell (RUC) that is the smallest possible geometry for regular honeycombs, but with a discrete layered cell wall, and (2) a tissue RVE based on microtomized cross-sections at different positions within a characteristic growth ring. For this case, the ESL representation is required, capturing the effect of microstructural disorder at the expense of not being able to spatially resolve relative viscoelastic deformations within the S layers of the cell wall. For spruce, the negligence of ray tissue, sap channels, or cell wall features like pits is sensible and acceptable, as it increases the predictive quality of micro-mechanical models, as larger models become feasible. 
\subsubsection{The Y-shaped tissue RUC}\label{subsubsec2-3-1}
\begin{figure}[htb]
  \centering{\includegraphics[width=0.7\linewidth]{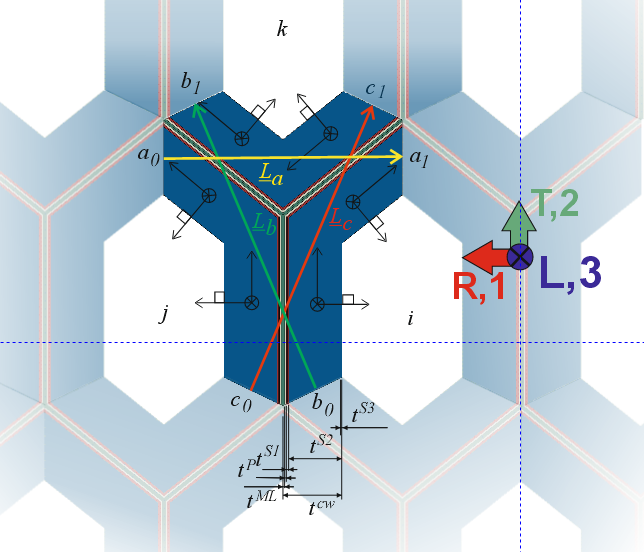}}
  \caption{\label{fig:y-rve} Irreducible Y-shaped RUC for LW with boundary conditions, the global material coordinate system, and local cell wall coordinate systems.}
\end{figure}

One of the simplest representations of the cellular structure of softwood is provided by a regular hexagonal grid \citep{gibson_cellular_1997,persson_micromechanical_2000}. Its behavior can be determined on an irreducible Y-shaped RUC, formed by segments of three neighboring tracheids, as shown in Fig.~\ref{fig:y-rve}. Only four parameters characterize the RUC geometry, namely the lengths $l$ and $h$ representing the radial and tangential size of the tracheids, respectively, the shape angle $\theta$, and the cell wall thickness $t_{cw}$ (see Fig.~\ref{fig:1new}). The RUC extends in the longitudinal direction with unit thickness. The average geometric properties of the tracheids ($l$, $h$, and $\theta$) were measured from images of cross sections. For each modeled tissue (EW, TW, and LW), we use a representative $t_{cw}$ obtained from the image-based approach described in \citet{ferrara_wittel_densification_2025} by averaging all measured cell-wall thicknesses, i.e. $t_{cw}^{EW}$, $t_{cw}^{TW}$, and $t_{cw}^{LW}$. Note that all tissue images were acquired from the same log of Norway spruce used in our previous experimental micro- and macro-scale campaigns \citep{ferrara_micro-mechanical_2024,ferrara_tensile_2025,maas_overview_2025,maas_comprehensive_2025}. The applied geometrical values of the Y-shaped RUCs are tabulated in Tab.~\ref{tab:cellgeom}.
\begin{table}[htb]
\centering
\caption{Geometrical parameters of the Y-shaped RUCs for EW, TW, and LW.}\label{tab:cellgeom}
\begin{tabular}{c| c c c c c}
    & \textbf{$l$} & \textbf{$h$} & \textbf{$\theta$} & \textbf{$t_{cw}$} & \textbf{$t_{S2}$} \\
    & ($\mu$m) & ($\mu$m) & (\SI{}{\degree}) & ($\mu$m) & ($\mu$m) \\ 
\hline
EW$\quad$  & 24.2 & 22   & 21.8 & 2.55 & 1.915 \\
TW$\quad$  & 21.0 & 22   & 25.3 & 3.85 & 3.215 \\
LW$\quad$  & 15.4 & 22   & 35.8 & 6.00 & 5.365 \\
\end{tabular}
\end{table}

Each tracheid in the RUC consists of six layers (ML, P, S1+, S1-, S2, S3), with material properties defined in Sec.~\ref{subsubsec2-2-1} and summarized in Tab.~\ref{tab:1} for $\omega$=\SI{12}{\percent}. Since each layer’s engineering constants are defined in its layer coordinate system, a distinct discrete material orientation is assigned to each layer that is defined by one axis being aligned with the micro fibrils of the respective layer and one being normal to the lumen surface. 20-node quadratic brick elements with reduced integration (C3D20R) are used for mesh generation and calculation, with the two cases of geometric nonlinearities being disabled and enabled. 

To function as a truly repetitive unit cell, the boundary edges must be coupled so that the displacement field remains continuous and periodic across adjacent RUCs while the forces are anti-periodic \citep{persson_micromechanical_2000}. Periodic boundary conditions (PBCs) are imposed along the directions of the lattice vectors $\mathbf{L}_a$, $\mathbf{L}_b$, and $\mathbf{L}_c$, which define the in-plane translational symmetry of the RUC (see Fig.~\ref{fig:y-rve}). The same principle applies in the longitudinal direction, where the lattice vector $\mathbf{L}_d$ couples the front and back faces of the unit cell. The vectors $\mathbf{L}_l$, with $l = a, b, c, d$, connect corresponding nodes on opposite boundaries: $a_0 \rightarrow a_1$, $b_0 \rightarrow b_1$, $c_0 \rightarrow c_1$, and $d_0 \rightarrow d_1$. For each pair of boundary edges, the following equations must be satisfied:
\begin{align}
u_1^{0} - u_1^{1} + L_{l,1} \, {\overline{\varepsilon}}_{11} + L_{l,2} \, {\overline{\varepsilon}}_{12} + L_{l,3} \, {\overline{\varepsilon}}_{13} &= 0 \label{eq:pbc1}\\
u_2^{0} - u_2^{1} + L_{l,1} \, {\overline{\varepsilon}}_{12} + L_{l,2} \, {\overline{\varepsilon}}_{22} + L_{l,3} \, {\overline{\varepsilon}}_{23} &= 0 \label{eq:pbc2}\\
u_3^{0} - u_3^{1} + L_{l,1} \, {\overline{\varepsilon}}_{13} + L_{l,2} \, {\overline{\varepsilon}}_{23} + L_{l,3} \, {\overline{\varepsilon}}_{33} &= 0 \label{eq:pbc3}
\end{align}
where $u_1, u_2, u_3$ are the components of the displacement vector $\mathbf{u}$; the superscripts $0$ and $1$ indicate corresponding nodes on opposite edges of the RUC; $L_{l,1}, L_{l,2}, L_{l,3}$ are the components of the lattice vector $\mathbf{L}_l$; and $\overline{\varepsilon}_{jk}$ are the components of the strain vector $\overline{\boldsymbol{\varepsilon}}$ representing the average deformation over the entire RUC.

To compute the equivalent orthotopic compliance tensor $\mathbf{C}^{-1,RUC}$ of the RUC, a master‐node technique similar to \citet{Rafsanjani_etal_2012} is employed. Each pair of periodic edges is coupled to reference points (RPs) whose displacements enforce the PBCs (Eqs.~\ref{eq:pbc1}–\ref{eq:pbc3}), since they impose the average strains $\overline{\varepsilon}_{jk}(t)$. Nine RPs are introduced: three for the axial strain components and six for the shear components (one for each related shear plane). In this way, $\mathbf{C}^{-1,RUC}$ can be evaluated at any time $t$ from the relation between the forces on the RPs and their resulting displacements, from which the average stresses $\overline{\sigma}_{jk}(t)$ and strains $\overline{\varepsilon}_{jk}(t)$ are obtained. Each component is then calculated as
\begin{equation}\label{eq:rvecomp}
\begin{gathered}
C_{AB}^{-1}(t) = \frac{\overline{\varepsilon}_{jk}(t)}{\overline{\sigma}_{jk}(t)} \ \text{for} \ A = B, \ \text{and} \
C_{AB}^{-1}(t) = \frac{\overline{\varepsilon}_{jj}(t)\;-\;C_{AA}^{-1}(t)\,\overline{\varepsilon}_{kk}(t)}{\overline{\sigma}_{kk}(t)} \ \text{for} \ A \neq B.
\end{gathered}
\end{equation}
Nine elementary loading cases are implemented: three uniaxial, three pure shear, and three biaxial to extract the diagonal and off‐diagonal compliance terms.
\subsubsection{The disordered tissue RVE}\label{subsubsec2-3-2}
\begin{figure}[htb]
  \centering{\includegraphics[width=0.9\linewidth]{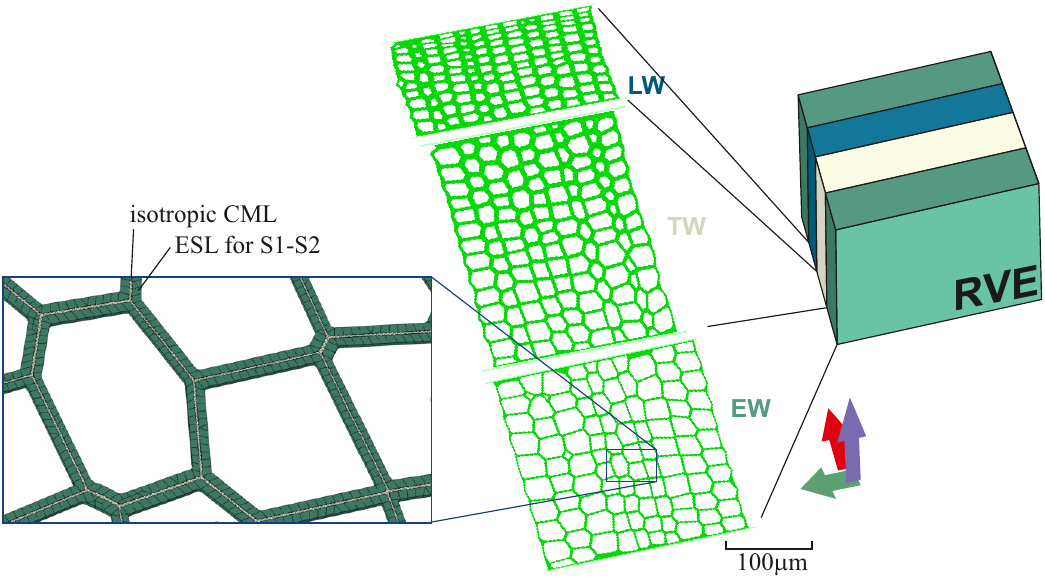}}
  \caption{\label{fig:rves} Models for the representative structures for early (EW), transition (TW), and late wood (LW). The magnification shows the mesh with C3D20R and C3D10 elements, and the local material coordinates with the orientation following the lumen wall.}
\end{figure}

To relax the strong assumption of structural regularity, one could generate a randomized structure from morphometric distributions of $h,l,\theta$, as demonstrated in \cite{mora_etal_2019}. Alternatively, one can directly work on microscopy images of RT-cross-sections and apply image analysis to build up representative tissue models that comprise a sufficient number of tracheids. The process from sample preparation, via image acquisition, enhancement, segmentation, and morphological analysis to the model construction in Abaqus via Python scripting is described in detail in \cite{ferrara_wittel_densification_2025}. To keep the model size within acceptable bounds for the inverse parameter identification, distinct early, transition, and late wood tissue RVS are constructed as shown in Fig.~\ref{fig:rves}. 

To reduce the computational effort, the cell walls are considered as an ESL (Sec.~\ref{subsubsec2-2-2}). As further simplification, the orthotropic compliance tensor of the S layers is calculated from an average S2 thickness, resulting in a single material per tracheid type (EW, TW, and LW) listed in Tab.~\ref{tab:cellgeom}. Note that these average thicknesses are used solely to compute material properties. At the same time, the FEM model has individual but uniform cell wall thickness per tracheid corresponding to the cross-sectional images. The resulting engineering constants, in the cell wall coordinate system at $\omega$=\SI{12}{\percent}, are reported in Tab.~\ref{tab:tissueprop2} alongside the average thicknesses used for their computation. Since the input material properties of each tracheid are defined in the cell wall coordinate system, a discrete material orientation is assigned to each tracheid that is defined by the rotation of the cell wall coordinate system around the L-direction in such a way that the transverse material orientation always lies on the inner lumen surface. At the same time, the other one is normal to it.
\begin{table}[htbp]
\centering
\caption{Engineering constants of compound middle lamella (CML) and equivalent single-layer (S) for early (EW), transition (TW), and late wood (LW) in GPa at $\omega$=\SI{12}{\percent}. The values are expressed in the cell wall coordinate system (1 = longitudinal direction (L); 2 = transverse direction ($\parallel$ wall); 3 = normal direction ($\perp$ wall)).}
\label{tab:tissueprop2}
\begin{tabular}{l|cccccccccc}
 & thick. & $E_1$ & $E_2$ & $E_3$ & $G_{23}$ & $G_{31}$ & $G_{12}$ & $\nu_{12}$ & $\nu_{13}$ & $\nu_{23}$ \\
 & ($\mu$m)  & \multicolumn{9}{c}{(\SI{}{\giga\pascal} / -)} \\
\hline
CML       & 0.35 & 3.41  & 3.41  & 3.41  & 1.27 & 1.27 & 1.27 & 0.35 & 0.35 & 0.35 \\
S$_{EW}$  & 2.20 & 49.39 & 7.89  & 4.82  & 1.62 & 1.65 & 6.15 & 0.70 & 0.17 & 0.35 \\
S$_{TW}$  & 3.50 & 51.51 & 6.63  & 4.80  & 1.64 & 1.67 & 5.98 & 0.81 & 0.14 & 0.35 \\
S$_{LW}$  & 5.65 & 52.76 & 5.81  & 4.78  & 1.65 & 1.69 & 5.87 & 0.90 & 0.10 & 0.35 \\
\end{tabular}
\end{table}

The S layers of the tracheids are discretized with C3D20R quadratic brick elements, while the matrix is represented by a combination of brick and tetrahedral elements (C3D10) for geometrical reasons (see Fig.~\ref{fig:rves}). Identical to the Y-shaped RUC (see Sec.~\ref{subsubsec2-3-1}), nine elementary loading cases are implemented (three uniaxial, three pure shear, and three biaxial) to populate the equivalent orthotropic compliance tensor $\mathbf{C}^{-1,RVE}$ of the tissue RVE for EW, TW, and LW.
\subsection{The growth ring RVE}\label{subsec2-4}
To calculate the macroscopic behavior, a growth ring RVE is defined. It comprises one growth ring discretized into three different materials, namely EW, TW, and LW (Fig.~\ref{fig:rves}). The average growth ring width $t_{gr}$ measured as \SI{1.4}{\milli\meter} for our tree is split in the ratio LW:TW:EW = 3:4:7. Note that the LW-EW interface of the RVE is not located at the RVE boundary but astray from the edge to be able to capture the important deformations at the growth ring interface correctly. The material properties of each tissue for the UMat can be determined either from the Y-shaped RUC (Sec.~\ref{subsubsec2-3-1}) or the tissue RVE (Sec.~\ref{subsubsec2-3-2}). Elastic properties are obtained by applying the appropriate boundary conditions without activated creep. In contrast, the viscoelastic properties are extracted from the creep compliances $J_{c,AB}(t)$, which are fitted to derive the proportional factors $\gamma_{i}^{ve, RVE/RUC}$ (Eq.~\ref{eq:pronyser2}) for each tissue type. A mesh of quadratic brick elements with reduced integration (C3D20R) is applied, which is sufficiently fine to minimize discretization effects. Identical to the other RCUs and RVEs, nine distinct loading cases are applied, and from the calculated response, the orthotropic compliance tensor $\mathbf{C}^{-1,GR}$ is populated at any time $t$. The elastic properties $\mathbf{C}^{-1,GR}_{0}$ are determined \textit{a priori}, whereas the creep compliance $\mathbf{J}_{c}^{GR}(t)$ is used in the surrogate model for identifying the proportionality parameter set for the viscoelastic behavior of the cell wall and its layers (see Sec.~\ref{subsec2-5}).
\subsection{Inverse parameter identification procedure}\label{subsec2-5}
As depicted in Secs.~\ref{subsec2-2}-\ref{subsec2-4} and Fig.\ref{fig:1new}, we sequentially calculate the elastic compliance tensor of the cell wall layer, then the one of the cell wall, the one of the different tissue types and finally with those tensors the one of the growth ring RVE that result in the calculated macroscopic response. The same procedure is made for the viscoelastic compliance tensor with the unknown proportionality factor set $\gamma_i^{ve}$ of the cell wall layers. By calculating one RVE after another, we ultimately obtain the macroscopic creep response. Since all elastic parameters are taken from literature and model topologies are fixed, we only need to identify the appropriate $\gamma_i^{ve}$-set in an inverse parameter identification procedure. To judge the validity of a $\gamma_i^{ve}$-set, we define a multi-criteria objective function that compares the creep response of the growth ring model with the experimental measurements in \citep{maas_comprehensive_2025}. The scalar multi-criteria objective function $f_{obj}$ is thus composed of \num{9} indicators $f_{AB}$, each weighted by $w_{AB}$, to quantify the agreement between simulated and experimental creep for each $AB$-component of the compliance tensors, namely
\begin{equation}\label{eq:objective}
  f_{\rm obj}
  = \sum_{\substack{AB}}
    w_{AB}
    \,
    \frac{
      \bigl\|
        (J_{c,AB}^{GR}(t)/C_{0,AB}^{-1,GR})^{\rm sim}
        -
        (J_{c,AB}^{GR}(t)/C_{0,AB}^{-1,GR})^{\rm exp}
      \bigr\|
    }{
      \bigl\|
        (J_{c,AB}^{GR}(t)/C_{0,AB}^{-1,GR})^{\rm exp}
      \bigr\|
    } \,.
\end{equation}
The overall aim of the optimization is to minimize the discrepancy between the experimentally determined \citep{maas_comprehensive_2025} and the simulated normalized creep response for various orientations of the growth ring, namely $\min(f_{\rm obj}(\gamma_i^{ve}))$ \citep{Gritzmann2013}. 

As simulations with the rheological tissue models and the successive growth ring model are computationally expensive, gradient-based optimization algorithms are not feasible. A single evaluation of the objective function requires, for a given parameter set, the generation of multiple FEM systems, their calculation and evaluation using Abaqus scripting. FEM simulations are performed in Abaqus on the isolated tissues (EW, TW, LW) to compute the respective creep compliance tensor $\mathbf{J}_{c}(t)$ which then provides the viscoelastic inputs for the growth ring simulations. The GR outputs are post-processed into normalized creep compliance curves $J_{c,AB}^{GR}(t)/C^{-1,GR}_{0,AB}$ for each $AB$-component, which serve as input for the evaluation of the objective function. Considering the nature of the problem, surrogate models \citep{Booker1999} that only approximate the objective function can be used to guide the search for a near-optimal solution within a reasonable number of function evaluations. The idea is to estimate the minimal value of $f_{obj}$ for an unknown $\gamma_i^{ve}$-parameter set, called design, based on previous evaluations. The locations in the parameter domain are used to fit a surrogate surface. The minima of the surrogate model, along with new guesses, define the new designs for the next iteration of the optimization. 

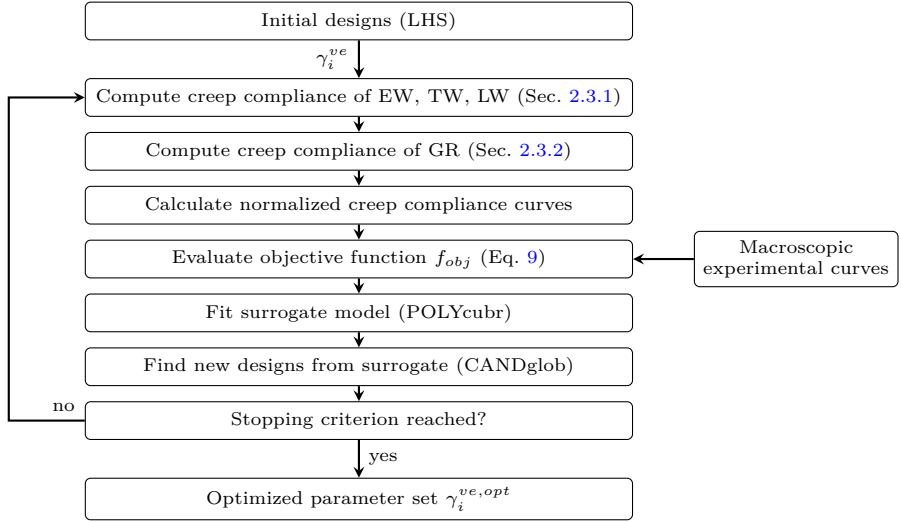
\begin{figure}[htb]
\begin{center}
\begin{tikzpicture}[node distance=0.2cm, auto]
  \node (init)[block]{Initial designs (LHS)};
  \node (abaqus1)[block, below=0.5cm of init]{Compute creep compliance of EW, TW, LW (Sec.~\ref{subsubsec2-3-1})};
  \node (abaqus2)[block, below=of abaqus1]{Compute creep compliance of GR (Sec.~\ref{subsubsec2-3-2})};
  \node (curves)[block, below=of abaqus2]{Calculate normalized creep compliance curves};
  \node (objf)[block, below=of curves]{Evaluate objective function $f_{obj}$ (Eq.~\ref{eq:objective})};
  \node (fit)[block, below=of objf]{Fit surrogate model (POLYcubr)};
  \node (newdesigns)[block, below=of fit]{Find new designs from surrogate (CANDglob)};
  \node (stop)[block, below=of newdesigns]{Stopping criterion reached?};
  \node (save)[block, below=0.5cm of stop]{Optimized parameter set $\gamma_i^{ve, opt}$};
   \node (macro) [block, minimum width=2cm, right=0.8cm of objf, align=center]{Macroscopic\\ experimental curves};
  \draw [arrow] (init) -- (abaqus1) node[midway,left,font=\footnotesize]{$\gamma_i^{ve}$};
  \draw [arrow] (abaqus1) -- (abaqus2);
  \draw [arrow] (abaqus2) -- (curves);
  \draw [arrow] (curves) -- (objf);
  \draw [arrow] (objf) -- (fit);
  \draw [arrow] (fit) -- (newdesigns);
  \draw [arrow] (newdesigns) -- (stop);
  \draw [arrow] (stop) -- (save) node[midway,right,font=\footnotesize] {yes};
  \coordinate (aux1) at ($(stop.west)+(-1cm,0)$);
  \coordinate (aux2) at (aux1 |- abaqus1.west);
  \draw [arrow]
    (stop.west) node[above left,font=\footnotesize] {no} -- (aux1) -- (aux2) -- (abaqus1.west);
  \draw [arrow] (macro) -- (objf);
\end{tikzpicture}
\end{center}
\caption{Optimization workflow using MATSuMoTo surrogate modeling.}\label{Matsumoto}
\end{figure}

In this work, MATSuMoTo (\textbf{MAT}LAB \textbf{Su}rrogate \textbf{Mo}delling \textbf{To}olbox) by \citet{Mueller1,Mueller4} was employed, which allows for an easy evaluation of different surrogate models. The overall optimization workflow is illustrated in Fig.~\ref{Matsumoto}. The initial set of designs $\gamma_i^{ve}$ for the cell wall material scale is generated using Latin Hypercube Sampling (LHS) and evaluated to fit the first surrogate model using the Polycube Regression method (POLYcubr). New parameter sets are then chosen using the CANDglob strategy, which combines randomized local perturbations of the best solution with uniform sampling across the entire parameter space to avoid convergence to local minima. MatSuMoTo proposes these candidates, which are then used to generate FEM systems for evaluating a new objective function. Iterations continue until no significant improvement is observed over several successive runs or until a maximum number of iterations is reached, as in this study.

As is common for multi-variate optimization problems, auxiliary constraints are needed to avoid unphysical values, e.g. $\gamma_i^{ve}>0$ for $i=1\ldots 4$. Once a set of parameters $\gamma_i^{ve, opt}$ is found for the cell wall material scale that yields satisfying agreement of the viscoelastic behavior of the two-order larger scale model with respect to a set of experiments, the resulting framework could explain the viscoelastic behavior, such as axial, shear, Poisson's ratios or assymmetries, for any loading case.

\section{Results and discussion}\label{sec3}
The multi-scale approach enables the analysis of both the instantaneous (Sec.~\ref{subsec3-1}) and rheonomous (Sec.~\ref{subsec3-2}) material responses at the tissue as well as the macroscopic growth ring scale. First, the elastic behavior of the EW, TW, and LW tissue for the regular Y-shaped RUC and the disordered tissue RVE is described and compared to literature findings (Sec.~\ref{subsec3-1-1}). The resulting tissue behavior is incorporated into the growth ring model to predict the macroscopic response, which is then compared to literature values (Sec.~\ref{subsec3-1-2}). Second, the outcome from the surrogate model optimization is presented, along with the resulting rheonomous orthotropic creep, with a focus on the effects of geometric nonlinearity and structural disorder on creep.

\subsection{Orthotropic elastic behavior}\label{subsec3-1}
By applying the boundary conditions described in Sec.~\ref{subsubsec2-3-1} for the different elementary load cases while neglecting creep, we calculated the elastic orthotropic compliance tensors and the corresponding engineering constants for EW, TW, and LW from the Y-shaped RCUs and tissue RVEs. Tab.~\ref{tab:tissueprop1} summarizes the engineering constants at $\omega$=\SI{12}{\percent} for the isolated tissue and compares them with the mesoscale parameters measured for EW and LW reported in \citet{ferrara_micro-mechanical_2024}. Note that the results for the RUC are obtained with geometric non-linearity deactivated, while it is activated for the RVE. The combination of the orthotropic behavior of the tissues into the growth ring RVE (see Sec.~\ref{subsec2-4}) produces elastic properties which are directly comparable to macroscopic measurements reported in \citet{maas_overview_2025}. For completeness, the mesoscopic measurements for the growth ring \citep{ferrara_micro-mechanical_2024} are also included. For an improved comparison between the simulation results, the compliances of the tissues from both modeling approaches are shown in the principal planes in Fig.~\ref{fig:defbodyp_polar}a. The results for the RUC with geometric non-linearity (NL) are also included. In Fig.~\ref{fig:defbodyp_polar}b, instead, the results for the growth ring are shown and confronted with macroscopic measurements from \citet{maas_overview_2025} and literature data from \citet{hassani_etal_2015} and \citet{persson_micromechanical_2000} at $\omega$=\SI{12}{\percent}.

\begin{table}[htbp]
\centering
\caption{Engineering constants of the different tissue types in MPa at $\omega$=\SI{12}{\percent}, calculated from the Y-shaped RCUs and tissue RVEs, compared to mesoscale experimental data with * from \citet{ferrara_micro-mechanical_2024}, macroscale experimental data with ** from \citet{maas_overview_2025}, and literature data from \citet{hassani_etal_2015,persson_micromechanical_2000}. For the mesoscale shear moduli *, the first value refers to the planes TL, RL, RT, while the second refers to the corresponding related planes.}
\label{tab:tissueprop1}
\begin{tabularx}{\textwidth}{%
    >{\centering\arraybackslash}p{2cm}  
    >{\centering\arraybackslash}X|*{9}{>{\centering\arraybackslash}X}
}
 &  & $E_R$ & $E_T$ & $E_L$ & $G_{TL}$ & $G_{RL}$ & $G_{RT}$ & $\nu_{RT}$ & $\nu_{RL}$ & $\nu_{TL}$ \\
 & & \multicolumn{9}{c}{(MPa / -)} \\
\hline
\multirow{4}{*}{\makecell{Y-shaped RUCs\\$\rightarrow$ GR-RVE}}
  & EW   &  259  &   90   &   9936  &   556  &   799  &    35  & 1.411 & 0.017 & 0.005 \\
  & TW   &  603  &  414   &  16662  &  1053  &  1164  &   125  & 0.791 & 0.026 & 0.016 \\
  & LW   &  1383 & 2396   &  29950  &  2489  &  1592  &   482  & 0.278 & 0.027 & 0.059 \\
  & GR   &  733  &  658   &  15825  &  1091  &   991  &    59  & 0.550 & 0.028 & 0.029 \\
\hline
\multirow{4}{*}{\makecell{Tissue RVEs\\$\rightarrow$ GR-RVE}} 
  & EW   &  357 &  322 &  7947 &  668 &  678 &   10 & 0.585 & 0.032 & 0.027 \\
  & TW   &  700 & 1036 & 16490 & 1305 & 1270 &   71 & 0.334 & 0.030 & 0.043 \\
  & LW   & 1619 & 2111 & 28343 & 2157 & 2078 &  233 & 0.258 & 0.040 & 0.053 \\
  & GR   &  602 &  899 & 14419 & 1159 &  935 &   18 & 0.315 & 0.029 & 0.042 \\
\hline
\multirow{4}{*}{} 
  & EW*  &       &    84   &   4967  & 50--244 &         &        &       &       &       \\
  & LW*  &       &   727   &  13897  & 307--489&         &        &       &       &       \\
  & GR*  &   660 &   306   &   6203  &         &195--322 & 12--14  &       &       &       \\
  & GR** &   922 &   671   &  14622  &  1476   &  1282   &   55   &0.420  &0.034  &0.030  \\
\hline
  \multicolumn{2}{l|}{\citet{hassani_etal_2015}} & 822 & 423 & 12002 & 745 & 629 & 43 & 0.598 & 0.028 & 0.019 \\
  \multicolumn{2}{l|}{\citet{persson_micromechanical_2000}} & 987 & 455 & 11160 & 681 & 751 & 9.68 & 0.147 & 0.039 & 0.018 \\
\end{tabularx}
\end{table}


\subsubsection{Orthotropic behavior of the tissues}\label{subsec3-1-1}
\begin{figure}[htb]
  \centering{\includegraphics[width=1\linewidth]{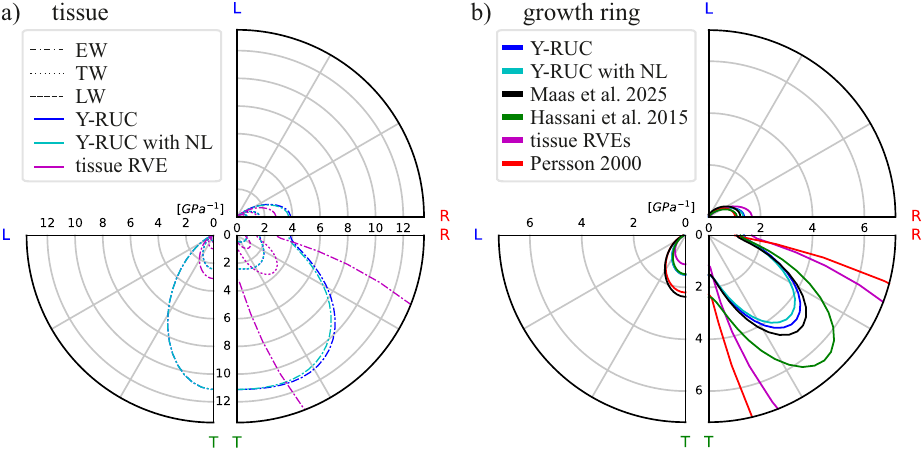}}
  \caption{\label{fig:defbodyp_polar} Compliances of the different RUCs and RVEs (a) and the respective macroscopic behavior (b) in the material symmetry planes compared to experimental data from \cite{maas_overview_2025} and literature data from \citet{hassani_etal_2015,persson_micromechanical_2000} at $\omega$=\SI{12}{\percent}.}
\end{figure}

The \textbf{compliances of the RUC with and without geometric non-linearity} (see Fig.~\ref{fig:defbodyp_polar}a) are obtained under the same applied loads for each elementary case. The curves are nearly identical, indicating that no significant non-linear deformations occur during these elastic calculations. Looking at the relationships between directional properties across tissues (Tab.~\ref{tab:tissueprop1}), all engineering constants of the Y-shaped RUC increase from EW to TW to LW as expected, while the Poisson's ratio $\nu_{RT}$ follows the inverse trend. Because T-direction deformation is bending-dominated, the tangential modulus $E_T$ remains lower than the radial modulus $E_R$ in EW and TW, while this trend reverses in LW due to significant morphologic changes in tracheid size and cell wall thickness. Although the predicted \(E_T\) for EW almost exactly matches experimental data, the LW value is noticeably overestimated, likely reflecting the thinner cell walls in real latewood and residual early- or transition-wood zones in the sampled LW slices. The ordering of the shear moduli $G_{RL}$ \textit{vs} $G_{TL}$ across the three tissues is the same as that for $E_R$ \textit{vs} $E_T$. Meanwhile, $G_{RT}$ remains substantially lower than both $G_{TL}$ and $G_{RL}$, since RT-plane shear deformation is dominated by bending of the cell walls. Finally, both $E_L$ and $G_{TL}$ predicted for EW and LW exceed the corresponding mesoscale measurements. Most likely, this difference occurs due to the different realized stress states between the experiment with plane stress and the simulation with generalized plane strain state \citep{ferrara_micro-mechanical_2024}. Note that optimizing elastic properties is not the scope of this study, as the primary focus is on creep behavior. Nevertheless, it is remarkable that the basic micromechanical assumptions for the layered cell wall model, combined with the tracheid geometry of the tested spruce, provide reasonable agreement with experimental data on the mesoscale.

The \textbf{engineering constants from the tissue RVEs} generally follow the same increasing progression from EW to LW as in the Y-shaped RCUs. In particular, the longitudinal modulus $E_L$ and the radial modulus $E_R$ from the two modeling approaches agree well across all tissue types, revealing that both geometries similarly capture the effects of material and cell walls orientation along with density in those directions. By contrast, the tangential modulus $E_T$ reveals a systematic discrepancy: for EW and TW tissue RVEs, $E_T$ is significantly higher than in the corresponding Y-shaped RCU, and even exceeds $E_R$, contrary to expectations. This could emerge from various simplifications, such as negligence of ray trisse, uniform cell wall thickness in R and T cell walls, or straightening of wall curvatures. One also observed a halving of $G_{RT}$ in each tissue RVE (even further for EW), whereas $G_{TL}$ and $G_{RL}$ remain close to the RUC. The Poisson’s ratios exhibit non-negligible discrepancies relative to the Y-shaped RCUs, with the largest deviations occurring when the T-direction is involved, particularly for EW. All these discrepancies are evident in Fig.~\ref{fig:defbodyp_polar}a, where the compliance curves for uniaxial longitudinal and radial deformation are close, and diverge markedly for the other deformation modes. Note that the described discrepancies are typical for wood tissue, but are not of relevance for further analysis and interpretation in this work.

\subsubsection{Observation of macroscopic behavior}\label{subsec3-1-2}
The composite GR-RVE behaves within expected ranges when compared to the macroscopic measurements \cite{maas_overview_2025}, in particular GR-RVE with the values from the Y-shaped RUC, as shown in Fig.~\ref{fig:defbodyp_polar}b and Tab.~\ref{tab:tissueprop1}. Note that the simulated radial modulus $E_R$ remains softer than the experimentally measured one, as the higher ondulation of load paths in R-direction softens the regular Y-shaped geometry. Interestingly, the tissue RVE–based model doesn't exhibit the same level of agreement in several aspects: $E_T$ is exceeding $E_R$, which can be traced back to the Poisson ratios of the tissue. Because the GR is composed of multiple layers that deform in parallel or in series depending on load orientation, each layer’s orthotropy affects the uniaxial stiffness. As a result, $E_R$ predicted by the tissue RVE ends up softer than both the experimental values and the Y-shaped RUC results, despite each individual tissue’s $E_R$ in the RVE exceeding its RUC counterpart. This could be an indicator of a not fully representative structural feature due to the limited size of the RVE. $G_{RT}$ is significantly lower than in the RUC and the experiments, reflecting the reduced $G_{RT}$ stiffness characteristic of the image-based model. This observation was also previously made in simulations by \cite{persson_micromechanical_2000} and is added to Fig.~\ref{fig:defbodyp_polar} for comparison. It is interesting to see that the typical butterfly-shaped deformation body of spruce, caused by the soft RT-shear direction (see one half of it in Fig.~\ref{fig:defbodyp_polar}), emerges mainly from the growth ring scale, and is not characteristic of the different Y-shaped RUCs.
\subsection{Orthotropic creep in wood}\label{subsec3-2}
Once the elastic compliance $\mathbf{C}_0^{-1}$ is determined, one can focus on the rheonomic compliances. As the $\gamma_i^{ve}$-sets of the cell wall layers are unknown, an optimization for the inverse parameter identification is performed. For the optimal $\gamma_i^{ve}$-sets determined for the regular Y-shaped RUC, the optimization progress is described first, before the creep behavior on the tissue scale, and finally, the macro-scale is discussed. In the same Section (Sec.~\ref{subsec3-2-1}), we explore the influence of geometric non-linearity on the creep behavior. Finally, we study the effect of disorder on the creep performance with the tissue RVEs in Sec.~\ref{subsec3-2-2}, all at $\omega=$\SI{12}{\percent} moisture content.
\subsubsection{Creep with regular hexagonal micro-structures}\label{subsec3-2-1}
\begin{figure}[htb]
  \centering{\includegraphics[width=0.8\linewidth]{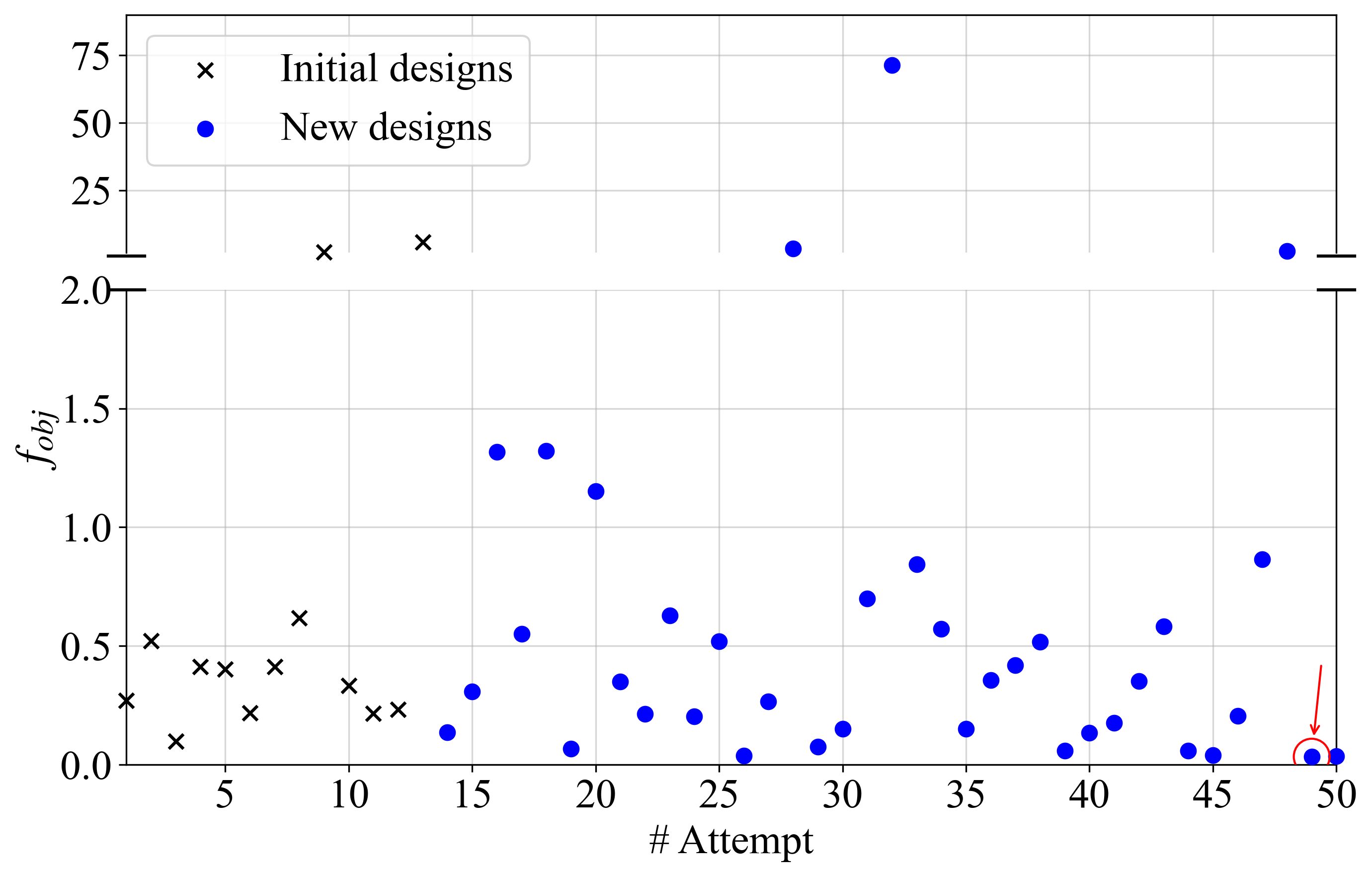}}
  \caption{\label{fig:iterations} Results of the objective function $f_{obj}$ for various designs throughout the surrogate optimization using MATSuMoTo. The red circle marks the minimum value reached.}
\end{figure}

The surrogate optimization with MATSuMoTo was performed with a maximum of \num{50} iterations as the stopping criterion. At each iteration, the creep compliances $J_{c,AB}(t)$ for EW, TW, LW, and then GR were computed under the nine elementary loading cases described in Sec.~\ref{subsubsec2-3-1}, applying a constant load for ar time of \SI{150}{\hour}. Fig.~\ref{fig:iterations} shows the development of the objective function $f_{obj}$ over all optimization attempts, with the red circle marking its minimum value $\min(f_{\rm obj}(\gamma_i^{ve}))$ which corresponds to the near-optimal parameters set $\gamma_i^{ve, opt}$.

\begin{figure}[htb]
  \centering{\includegraphics[width=\linewidth]{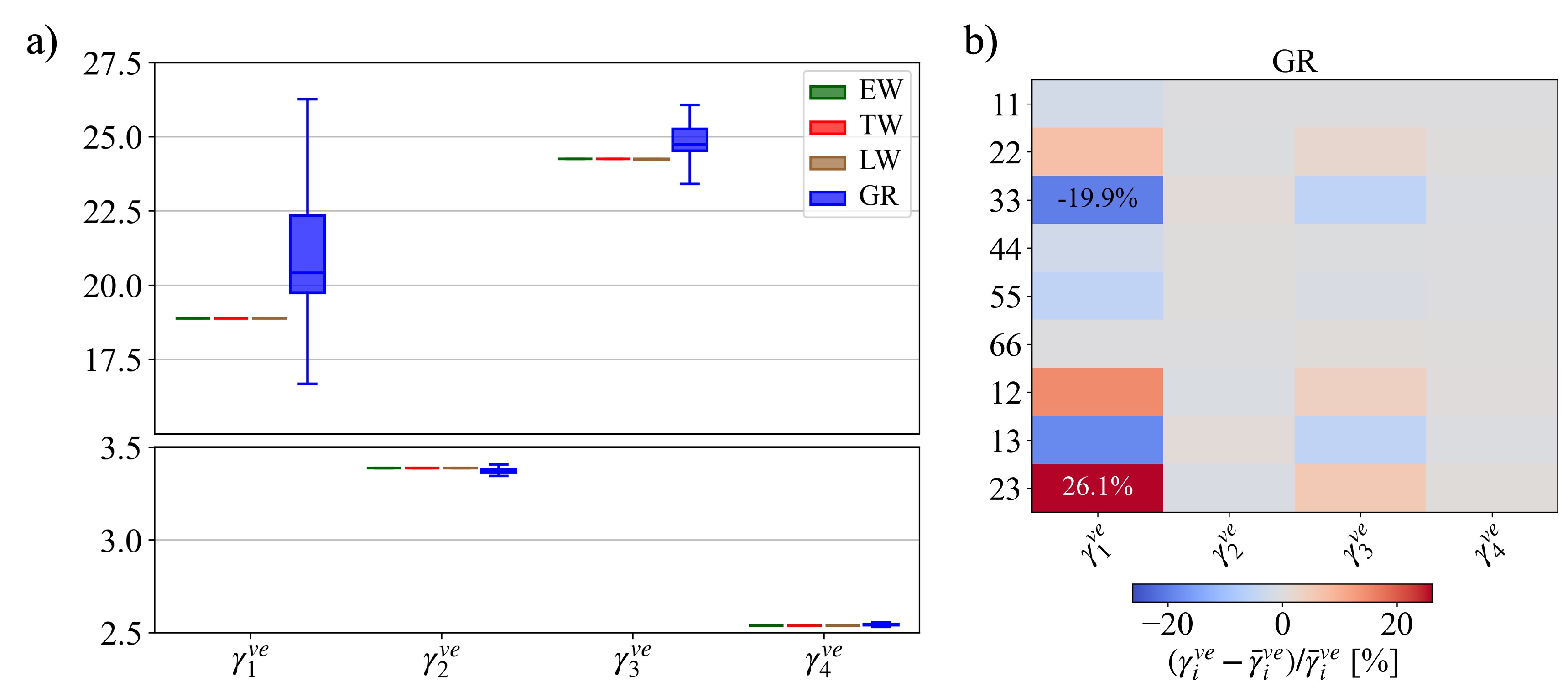}}
  \caption{\label{fig:gamma_distr} (a) Distributions of the proportionality factors $\gamma_{i,AB}^{ve,\mathrm{RUC}}$ for each Y-shaped RUC, EW, TW, and LW (without geometric non-linearity), and $\gamma_{i,AB}^{ve,\mathrm{GR}}$ for the growth ring across all $AB$-components, grouped by the $i^{th}$  element of the KV-chain. (b) Relative deviations of each $\gamma_{i,AB}^{ve}$ from its component-wise mean $\overline{\gamma}_i^{ve}$ for the growth ring.}
\end{figure}

The surrogate optimization was carried out on the Y-shaped RUC models with and without geometric non-linearity. The creep compliances $J_{c,AB}^{RUC}(t)$ for EW, TW, and LW are fitted with Eq.~\ref{eq:pronyser2} to extract the proportionality factors $\gamma_{i,AB}^{ve,\mathrm{RUC}}$. The near-optimal parameter set $\gamma_i^{ve,\mathrm{opt}}$ was found to be $[18.880, 3.386, 24.257, 2.537]$. Fig.~\ref{fig:gamma_distr}a shows the distributions of $\gamma_{i,AB}^{ve,\mathrm{RUC}}$ across all $AB$-components, grouped by the $i^{th}$  element of the KV-chain. Remarkably, for each $\gamma_{i}^{ve}$, the values collapse onto a single proportionality factor, i.e. they are essentially identical across all components and all three RUCs, and, more than this, they match the input set $\gamma_i^{ve,\mathrm{opt}}$. This implies that the equivalent creep behavior of each Y-shaped RUC can be described by a single set of average $\overline{\gamma}_i^{ve,\mathrm{RUC}}$. Although each layer has distinct elastic compliance and local material orientation, leading to different creep responses in the global $R-T-L$ coordinate system, their combined equivalent orthotropic cell wall behavior preserves the same elastic-to-viscoelastic proportionality as the layers. This consistency holds for all RUCs, proving the analogy between elastic and viscoelastic deformation mechanisms holds, regardless of local orientation or topological differences, providing a common $\overline{\gamma}_i^{ve,\mathrm{RUC}}$-set.

\begin{figure}[htb]
  \centering{\includegraphics[width=\linewidth]{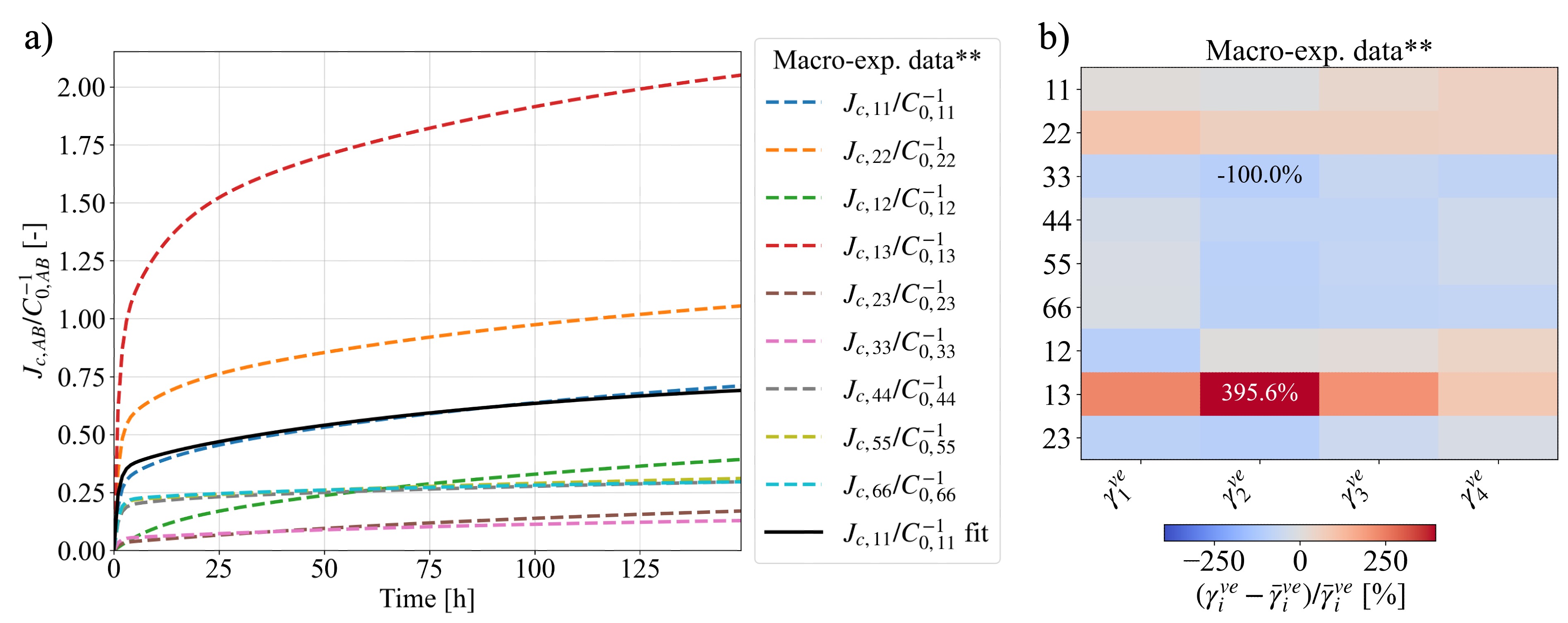}}
  \caption{\label{fig:macro_comp} (a) Normalized creep compliances from macroscale experimental data (**) reported in \citet{maas_overview_2025} compared with the normalized creep curve resulting from the surrogate optimization with MATSuMoTo. (b) Relative deviations of each $\gamma_{i,AB}^{ve}$ from its component-wise mean $\overline{\gamma}_i^{ve}$ for the growth ring.}
\end{figure}

With the common $\overline{\gamma}_i^{ve,\mathrm{RUC}}$-set, we can study the viscoelastic orthotropy of the Y-shape RUC-based growth ring. It exhibits a similar trend towards the collapse of proportionality factors. However, as shown in Fig.~\ref{fig:gamma_distr}a (blue boxes), the distributions of $\gamma_{i,AB}^{ve,\mathrm{GR}}$ are broader than those of the tissue RUCs, particularly for the short term first KV element ($\tau_1$ = \SI{0.1}{\hour}). Fig.~\ref{fig:gamma_distr}b shows the magnitude of the relative deviations of each $\gamma_{i,AB}^{ve}$ from its component-wise mean $\overline{\gamma}_i^{ve}$. Clearly, these variations can be considered as minor, justifying the use of a single averaged set of $\overline{\gamma}_{i}^{ve,\mathrm{GR}}$ to describe the creep behavior of the growth ring. The calculation confirms the assumption that the identical proportionality factors hold even across two scales. As shown in Fig.~\ref{fig:macro_comp}a, the experimental normalized creep compliances \cite{maas_comprehensive_2025} used for the MATSuMoTo optimization spread over a considerable range. In contrast, the calculated ones collapse into a single master curve. The relative deviations of each $\gamma_{i,AB}^{ve}$ from its component-wise mean $\overline{\gamma}_i^{ve}$ of the actual macroscopic measurements (Fig.\ref{fig:macro_comp}b) are visualized in the same way as for the GR-RVE. Clearly, the simulations do not reproduce directional dependencies. Note that, therefore, the objective function $f_{obj}$ can work with a creep curve for a single direction, e.g., $J_{c,11}/C_{0,11}^{-1}$.

\begin{figure}[htb]
  \centering{\includegraphics[width=\linewidth]{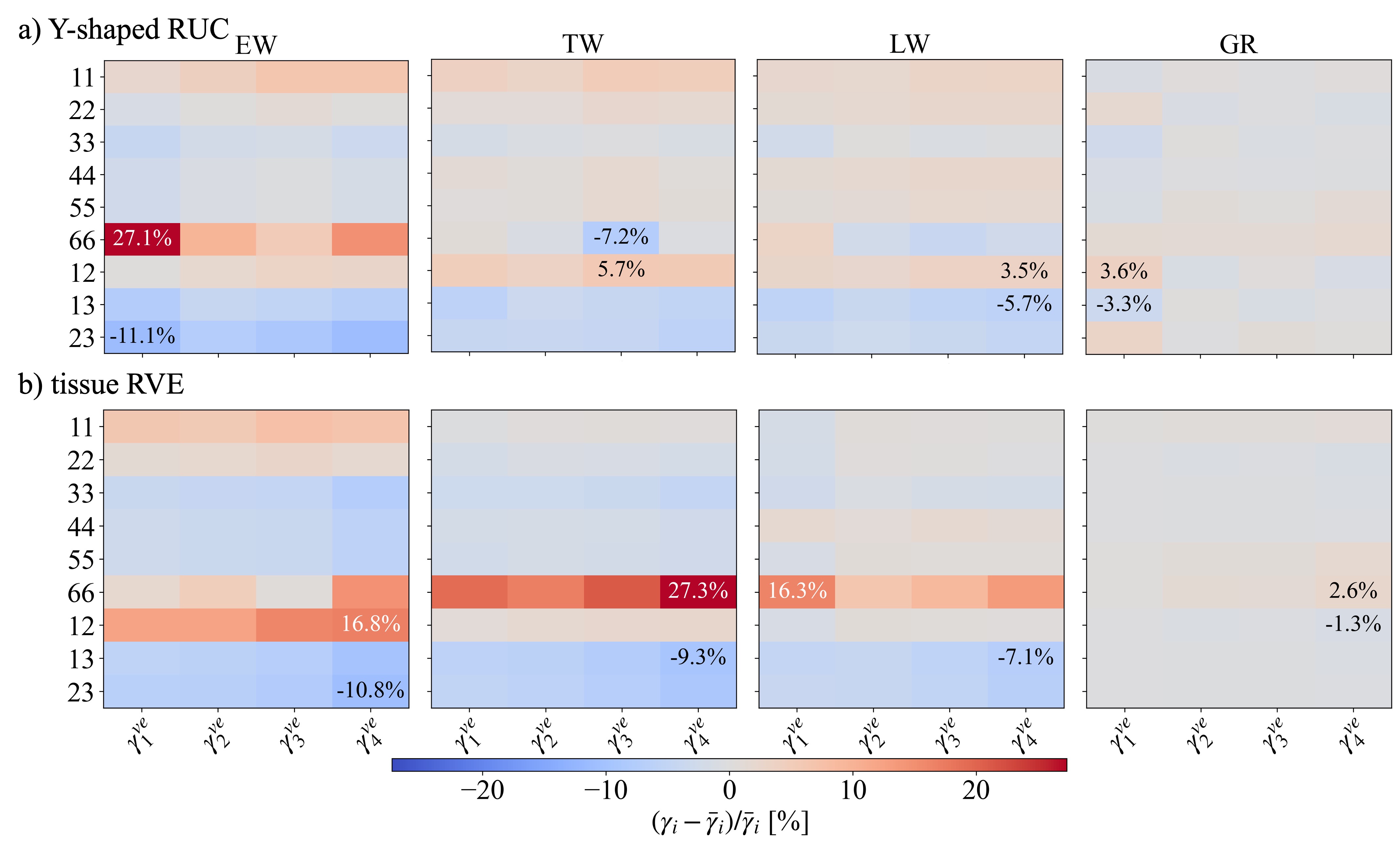}}
  \caption{\label{fig:colormaps} Relative deviations of each $\gamma_{i,AB}^{ve}$ from its component-wise mean $\overline{\gamma}_i^{ve}$ for Y-shaped RUC based model with geometric non-linearity (a) and tissue RVE based models (b).}
\end{figure}

One reason that directional dependencies did not emerge could be the role of geometric non-linearity. For simplicity, a single round of simulations with the same $\gamma_i^{ve,\mathrm{opt}}$-set, without any further optimization but with geometric non-linearity for the RUC is performed. Fig.~\ref{fig:colormaps}a gives the relative deviations of each $\gamma_{i,AB}^{ve}$ from its component-wise mean $\overline{\gamma}_i^{ve}$ for all tissue RUCs as well as the GR. While moderate discrepancies emerge at the tissue level, they cancel out and become insignificant for the growth ring. Hence, even with geometric non-linearity, the equivalent creep behavior of each Y-shaped RUC and the derived GR can be described by a common set of average proportional factors ${\overline{\gamma}_i^{ve}}$ which is nearly identical to the input $\gamma_i^{ve,\mathrm{opt}}$. This proves that geometric non-linearity alone cannot account for the observed directional dependencies of wood creep behavior.
\subsubsection{The role of disorder on creep}\label{subsec3-2-2}
Micro-structural disorder could be another reason for directional dependencies. To evaluate the impact of structural disorder on creep behavior, we performed a single round of simulations using the previous $\gamma_i^{ve,\mathrm{opt}}$ set, analogous to the case of geometric nonlinearity on the RUC. Fig.~\ref{fig:colormaps}b visualizes the relative deviations of each $\gamma_{i,AB}^{ve}$ from its component-wise mean $\overline{\gamma}_i^{ve}$ for all tissue RVEs and the GR. Again, the largest discrepancies emerge at the tissue level. In each model, the most considerable deviations are observed for shear in the RT plane, though these remain small compared with the spread in experimental tests (Fig~\ref{fig:macro_comp}b). When combined into the GR-RVE, they again more or less disappear, showing that the same set of ${\overline{\gamma}_i^{ve}}$ emerges across all tissues and on the macroscale. We therefore conclude that the real topology also cannot explain the directional dependencies of creep in wood.
\section{Conclusions and outlook}\label{sec4}
We demonstrated that all numerically determined creep compliance curves collapse to a certain extent towards a single set of proportionality factors. Consequently, the experimentally observed directional dependency of the viscoelastic response \cite{ferrara_tensile_2025,maas_comprehensive_2025}, observed in experiments, does not emerge from topological details on the tissue scale. As a matter of fact, the viscoelastic compliance tensor remains proportional to the elastic one, independent of the cellular configuration, even when multiple scales are involved. One must therefore conclude that directional differences do not emerge from topology, but must relate to non-linear material behavior at stress concentration points. On the cellular scale, those are cell corners, where hinges can form. Since those vary, depending on the load, directional dependencies can arise. 

The described approach has the potential for improvement in various directions. Clearly, a non-linear material response at the level of the cell wall layers would add richness to the directional dependence of creep, as deformations of bent, layered cell walls are not homogeneous but localized in cell corners, forming hinges with localized deformations. In this work, only a single moisture content was considered. However, one can easily relax this constraint to account for variable moisture content, as moisture-dependent elasticity tensors are used for the constituents in the first place. Changing moisture contents typically go along with mechanosorptive creep components, which can be considered with the Kelvin-Voight element series as well, only in the moisture rate domain \cite{hassani_etal_2015}. However, adding this complexity further complicates the parameter identification process. Furthermore, we assumed isotropy for the fixed characteristic times of the respective KV-elements. For practical reasons, this assumption is reasonable. However, its interpretation is strongly dependent on the length scale and is hindered by a lack of knowledge, as only speculative ideas on intra- or intercellular creep mechanisms exist \cite{engelund_tensile_2012}. Nevertheless, it is reasonable to assume one ruling mechanism, as suggested for slip-stick behavior in fiber bundles \cite{PhysRevE.109.044139}. Due to the microstructure of wood, any deformation mode would activate the exact mechanism in various locations and layers to varying degrees, justifying an isotropic characteristic time. 

\backmatter
\section*{Declarations}
\begin{itemize}
\item Acknowledgments: We acknowledge the Swiss National Science Foundation for funding this work under SNF grant 200021192186 "Creep behavior of wood on multiple scales". We also acknowledge Prof. Dr. Ingo Burgert for his valuable advice and ongoing support.
\item Funding: This work was funded by the Swiss National Science Foundation under SNF grant 200021192186 "Creep behavior of wood on multiple scales".
\item Conflict of interest: The authors state no conflict of interest.
\item Ethics approval and consent to participate: Not applicable.
\item Data availability: Upon request
\item Authors' contributions: Conception and methodology: FKW, AF; Formal analysis and investigation: AF.; Writing: original draft preparation, review, and editing: FKW, AF. Supervision: FKW.
\end{itemize}


\bibliographystyle{sn-basic.bst}
\bibliography{references}

\end{document}